\documentstyle[epsf,floats,multicol,prl,aps,eqsecnum]{revtex}
\newcommand{\rmt}{{\rm t}}
\newcommand{\rmd}{{\rm d}}
\newcommand{\rmnp}{{\rm p}}
\newcommand{\rmr}{{\rm r}}
\newcommand{\rmc}{{\rm c}}
\newcommand{\rmrt}{{\rm rt}}

\newcommand{\Det}{{\rm Det}}
\newcommand{\Tr}{{\rm Tr}}
\newcommand{\rmWZW}{{\rm WZW}}
\newcommand{\calH}{{\cal H}}
\newcommand{\calP}{{\cal P}}
\newcommand{\calT}{{\cal T}}
\newcommand{\calE}{{\cal E}}
\newcommand{\calL}{{\cal L}}
\newcommand{\calZ}{{\cal Z}}
\newcommand{\calD}{{\cal D}}
\newcommand{\calO}{{\cal O}}
\newcommand{\calF}{{\cal F}}
\newcommand{\tr}{{\rm tr}}

\newcommand{\tPsi}{{\widetilde\Psi}}
\newcommand{\tpsi}{{\widetilde\psi}}
\newcommand{\tchi}{{\widetilde\chi}}
\newcommand{\tb}{{\widetilde b}}

\newcommand{\cPsi}{\Psi^\circ}
\newcommand{\tcPsi}{\tPsi^\circ}

\newcommand{\slsh}[1]{{\not\! #1}}
\newcommand{\Wrt}{{W_\rmrt}}
\newcommand{\Wr}{{W_\rmr}
\newcommand{\Wt}{{W_\rmt}}}
\newcommand{\SpSp}{Sp($n)\times$Sp($n$) }
\begin{document}
\draft                                        
\preprint{}                                   
\title{ Phase diagram of disordered fermion model 
on two-dimensional square lattice with $\pi$-flux} 
\author{ Takahiro Fukui \cite{Email} }
 \address{
Department of Mathematical Sciences,
Ibaraki University, Mito 310-8512, Japan }
\date{\today}
\maketitle
\begin{abstract}
A fermion model
with random on-site potential defined on a
two-dimensional square lattice with $\pi$-flux is studied.
The continuum limit of the model
near the zero energy yields Dirac fermions
with random potentials specified by four independent 
coupling constants.
The basic symmetry of the model is time-reversal invariance.
Moreover, it turns out that the model has enhanced (chiral) symmetry
on several surfaces in the four-dimensional
space of the coupling constants.
It is shown that one of the surfaces with chiral symmetry
has \SpSp symmety whereas others have U($2n$) symmetry, both of which 
are broken to Sp($n$), and 
the fluctuation around a saddle point is described, respectively, by
Sp($n)_2$ WZW model and U($2n$)/Sp($n$) nonlinear sigma model.
Based on these results, we propose a phase diagram of the model.
\end{abstract}


\section{Introduction}\label{s:Int}

Anderson localization has attracted renewed
interest since a plenty of new universality classes were discoverd
\cite{Gad,Zir,AltZir}.
One is a system with particle-hole symmetry studied by Gade \cite{Gad}. 
Although generic disorderd systems in two dimensions are believed to be 
insulators \cite{AALR}, she showed that such a system has a
random critical point at the band center.
Typical examples studied so far are the random flux model 
\cite{Fur,AltSim,GLL,FabCas} 
and the random-hopping model with $\pi$-flux
\cite{HWKM,Fuk,GLL,FabCas}.
This class is often referred to as $A$III and $BD$I 
in the case with broken and
unbroken time reversal symmetry, respactively, which corresponds to the
chiral Gaussian unitary and chiral Gaussian orthogonal ensamble 
of the random matirx theory in the zero dimensional limit \cite{Zir}. 

Other examples of new universality classes are found in  
disordered $d$-wave superconductors described by
a mean-field Bogoliubov-deGennes Hamiltonian
\cite{NTW,ZHH,SFBN,BCSZd,SenFis,BSZ,Fukd,VisFis,FenKon,AHM,ZST,AHMZ,ASZ}.
Altland and Zirnbauer \cite{AltZir,Zir} classified them 
by two basic symmetries, i.e, spin-ratation and time-reversal into
four kinds of classes.
With (without) spin-rotation invariance,
disordered $d$-wave superconductors 
belong to class $C$ and $C$I ($D$ and $D$III), respectively, 
in the case with broken and unbroken time-reversal symmetry. 
The beta function of the renormalization group for
corresponding nonlinear sigma models 
in two dimensions tells that the classes $C$ and $C$I have localized
states only, whereas the classes $D$ and $D$III may have delocalized
states. What is particularly interesting is that some of them allow 
in the nonlinear sigma model description 
the topological term \cite{SenFis,BSZ,ASZ}
which plays a crucial role, for example,
 in quantum Hall transitions \cite{Pru},
or the Wess-Zumino-Witten (WZW) term \cite{Fukd,FenKon,ASZ}
which gives rise to a nontrivial strong-coupling fixed point.
Therefore, disordered $d$-wave superconductors are expected to have
quite rich phase diagram.

These developments can shed new light on the localization problems.
For example, the question whether the density of states of the 
random flux model diverges had been a long-standing problem.
Recognizing, however, that it belongs to the Gade's class of $A$III type, 
Furusaki concluded that the DOS actually diverges.
Another example is the disordered fermion model with $\pi$-flux
we will revisit in this paper.
This model was already studied by Fisher and Fradkin \cite{FisFra} in 1984. 
They showed that the effective field theory of the lattice model with 
random on-site potentials belongs to the universality class of
O($2n,2n$)/O($2n)\times$O($2n$) nonlinear sigma model,
concluding thereby that all states are localizaed.

A point here is that the continuum limit of the model 
includes random potentials specified by
four independent coupling constants.
Since Fisher and Fradkin studied only the case where all the coupling 
constants are equal,
we can expect richer phase diagram in the whole space of the
coupling constants.
Actually, although the generic points in this space
should flow into the localized phase mentioned-above, 
the model has enhanced symmetry in special cases and 
accordingly renormalization group flows in them are
toward different fixed points.
Another point in this paper is concerned with chiral symmetry.
It was shown that 
generic disordered $d$-wave superconductors in class $C$I,
described by Dirac fermions, yield no WZW term \cite{Fukd,ASZ} although
those Dirac fermions have chiral \SpSp symmetry.
The reason may be that chirality is not unique in the $d$-wave 
Dirac fermions,
since the lattice fermion gives the species doubling in the continuum
limit.  Here, chiral symmetry means that there exist a matrix which 
anticommutes with a Hamiltonian. 

In this paper, we investigate in detail the phase diagram of the disordered 
fermion model with $\pi$-flux by means of the fermionic replica method
to clarify these points.
We show that this model, described by double Dirac fermions,
has three kinds of chiral symmetry in special cases.
Interestingly, these different chiral symmetries 
lead to different universality classes.
To be concrete, an effective action of the Green functions 
has \SpSp symmetry in one of them and U$(2n)$ symmetry in the others, 
both of which are broken down to Sp($n$). 
It turns out that corresponding Goldstone mode can be described 
by Sp($n)_2$ WZW model and U($n$)/Sp($n$) nonlinear sigma model, respectively.

This paper is organized as follows. In the next section, we introduce
a lattice model, which is the same as Fisher and Fradkin studied, and derive 
Dirac Hamiltonian in the continuum limit. 
Although the most generic model has only time-reversal invariance, 
symmetry is enhanced on three surfaces in the space of the coupling
constants.
We focus our attention to one of them with
\SpSp symmetry (others have U($2n$) symmetry) 
and give calcualtions in detail.
Based on the continuum Hamiltonian, we derive an effective generating
functional of ensemble-averaged Green functions in Sec. \ref{s:RepMet}.
We show that \SpSp symmetry is spontaneously broken down to Sp($n$).
In Sec. \ref{s:ActGol}, we derive an effective action for the
Goldstone  mode, resulting in Sp($n)_2$ WZW model
(U($2n$)/Sp($n$) nonlinear sigma model for the others). 
In Sec. \ref{s:PhaDia}, we present the phase diagram. 
In Sec. \ref{s:Sum}, we give a summary and concluding remarks.

\section{Model}\label{s:Mod}

\subsection{A lattice model and its continuum limit}

We define a baisc lattice model 
on the square lattice in two dimensions.  For convenience,
we introduce similar notations used by Ludwig et al \cite{LFSG} 
in the following way. 
\begin{figure}[htb] 
\epsfxsize=70mm 
\centerline{\epsfbox{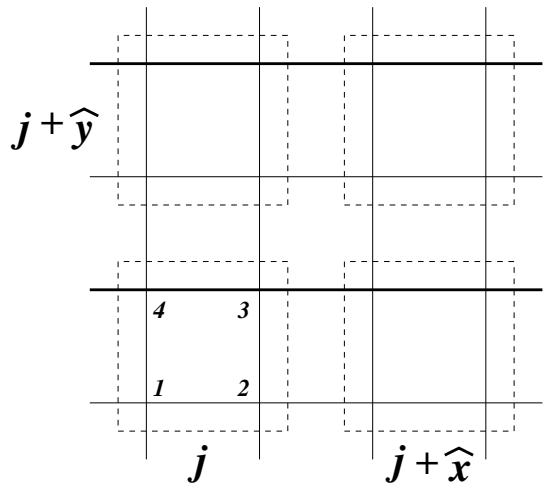}} 
\vspace{0.2cm}
\caption{The square lattice on which the model is defined.
Unit cells (plaquettes) are enclosed by the dotted-lines. Four sites 
on each plaquette are numbered counterclockwise.
Thick lines means that nearest neighbor
hopping is negative on them due to $\pi$ flux.}
\label{f:fig1}
\end{figure}
The square lattice in two dimensions is divided into unit cells,
each composed of a plaqutte enclosing four sites.
A plaquette is labled by a set of integers $j\equiv(j_x,j_y)$
and four sites on it are numbered 1,2,3 and 4 counterclockwise,
as illustrated in Fig. \ref{f:fig1}.
With such a notational convention,
a pure and disorder Hamiltonian 
denoted respectively as $H_\rmnp$ and $H_\rmd$ are defined by
\begin{eqnarray}
&&
H_\rmnp=\sum_j
\left(
 c_{1j}^\dagger c_{2j}+c_{2j}^\dagger c_{3j}
-c_{3j}^\dagger c_{4j}+c_{4j}^\dagger c_{1j}
+c_{2j}^\dagger c_{1j+\hat x}
-c_{3j}^\dagger c_{4j+\hat x}
+c_{3j}^\dagger c_{2j+\hat y}
+c_{4j}^\dagger c_{1j+\hat y}
+\mbox{h.c.}
\right) ,
\nonumber\\
&&
H_\rmd=\sum_j\sum_{a=1}^4v'_{aj}c_{aj}^\dagger c_{aj},
\label{LatHam}
\end{eqnarray}
where $\hat x\equiv(1,0)$, $\hat y\equiv(0,1)$, and
$v'_{a j}$ is random real number due to nonmagnetic impurities. 
Negative signs of the third and the sixth terms in $H_\rmnp$ are 
due to $\pi$ flux imposed.
Now let us construct an effective Hamiltonian near the band center
in the case of weak disorder. To this end, we first derive the continuum limit
of the pure model, and next we take into account the effects of disorder.

Fourier transformation 
$c_{a j}=\int\frac{d^2k}{(2\pi)^2}e^{ikj}c_a(k)$ 
for $a=1,...,4$ leads to 
\begin{eqnarray}
H_\rmnp=\int\frac{d^2k}{(2\pi)^2}
c^\dagger(k)
\left(
\begin{array}{cccc}
            &1+e^{-ik_x} &                &   1+e^{-ik_y}  \\
1+e^{ ik_x} &            &   1+e^{-ik_y}  &                \\
            &1+e^{ ik_y} &                & -(1+e^{ ik_x}) \\
1+e^{ ik_y} &            & -(1+e^{-ik_x}) &                
\end{array}
\right)
c(k),
\nonumber
\end{eqnarray}
where $c^\dagger(k)\equiv(c^\dagger_1(k),...,c^\dagger_4(k))$.
It is readily seen that the Hamiltonian gives zero-energies at 
$k=(\pi,\pi)$.
In the vicinity of them, the model can be effectively described
by Dirac fermions. To see this,
set $k=(\pi,\pi)+a_0p$ with the lattice constant $a_0$, 
expand the Hamiltonian to the first order of $p$, and
introduce a field operator in the continuum limit
$c_a((\pi,\pi)+a_0p)\equiv a_0^{-\frac{3}{2}}\psi_a(p)$ 
for $a=1,...,4$.
Then we have
\begin{eqnarray}
H_\rmnp\sim 
\int\frac{d^2p}{(2\pi)^2}
\psi^\dagger(p)
\left(
-\sigma_2\otimes1 p_x-\sigma_1\otimes\sigma_2 p_y
\right)
\psi(p) ,
\nonumber
\end{eqnarray}
where 
$\psi^\dagger=(\psi_1^\dagger,\psi_2^\dagger,
\psi_3^\dagger,\psi_4^\dagger)$.
Accordingly, effective Hamiltonian near the zero energy can be expressed 
by the following field theoretical Hamiltonian
\begin{eqnarray}
H_\rmnp=
\int d^2x 
\psi^\dagger(x) \calH_\rmnp \psi(x) , \quad
\calH_\rmnp\equiv \gamma_\mu i\partial_\mu ,
\label{ConPurHam}
\end{eqnarray}
where 
$x=a_0j$, 
$\psi(x)\equiv\int\frac{d^2p}{(2\pi)^2}e^{ipx}\psi(p)$, 
and $\gamma$ matrices are defined as
$\gamma_1=\sigma_2\otimes1$ and $\gamma_2=\sigma_1\otimes\sigma_2$.

By using the relation between the lattice operator 
$c_{a j}$ and the continuum field operator $\psi_a(x)$,
\begin{eqnarray}
c_{a j}
\sim
\int\frac{d^2(a_0p)}{(2\pi)^2}e^{i(\pi,\pi)j+ia_0pj}
a_0^{-\frac{3}{2}}\psi_a(p)
=a_0^\frac{1}{2}(-)^{\frac{x}{a_0}+\frac{y}{a_0}}\psi_a(x) ,
\nonumber
\end{eqnarray}
the continuum limit of the disorder potentials are readily derived;
$H_\rmd=\int d^2x \sum_{a=1}^4
\psi_a^\dagger(x) v'_a(x) \psi_a(x)$,
where $v'_a(x)$ is the slowly-varying part of $v'_{aj}$ defined as
$v'_{aj}\equiv a_0v'_a(x)$.
Taking suitable linear combination of $v'_a$, we can rewrite the 
random potentials as
\begin{eqnarray}
H_\rmd=\int d^2x \psi^\dagger(x)\calH_\rmd\psi(x),
\quad
\calH_\rmd=\sum_{\nu=1}^4v_\nu(x)\alpha_\nu ,
\label{ConDisHam}
\end{eqnarray}
where
$
\alpha_1=1\otimes\sigma_3,      
\alpha_2=\sigma_3\otimes\sigma_3,
\alpha_3=\sigma_3\otimes 1
$ and
$      
\alpha_4=1\otimes 1 
$.
We assume that the four disorder potentials obey Gaussian 
probability distribution
\begin{eqnarray}
P[v_\nu(x)]\propto e^{-\frac{2}{g_\nu}v_\nu^2(x)}, 
\quad \mbox{for}\quad \nu=1,...,4.
\label{ProDis}
\end{eqnarray}

It may be more convenient to switch into another basis via suitable 
rotation for $\psi$,
\begin{eqnarray}
&&
\gamma_1=\sigma_2\otimes1,        \quad
\gamma_2=\sigma_1\otimes\sigma_3, \quad 
\nonumber\\
&&
\alpha_1=1\otimes\sigma_2,        \quad
\alpha_2=\sigma_3\otimes\sigma_2, \quad
\alpha_3=\sigma_3\otimes 1,       \quad
\alpha_4=1\otimes1 .
\label{GamAlp}
\end{eqnarray}
In what follows, we refer to the space where these matrices live 
as $V$. It is distinguished from the replica space $W_\rmr$ 
introduced momentarily.
The total Hamiltonian we investigate in this paper is $H=H_\rmnp+H_\rmd$ 
in Eqs. (\ref{ConPurHam}) and (\ref{ConDisHam}) with (\ref{GamAlp}).
It is noted that this is the same model as
Fisher and Fradkin \cite{FisFra} studied in the case  $g_1=g_2=g_3=g_4$.
They showed that the effective field theory  
is O($2n,2n$)/O($2n)\times$O$(2n$) nonlinear sigma model
for the bosonic replicas
[or Sp($2n$)/Sp($n)\times$Sp($n$) nonlinear sigma model
for the fermionic ones], 
concluding thereby that all states are localized. 
According to Zirnbauer, this class is referred to as $A$II \cite{Zir}. 
Although a slight relaxation of the condition $g_1=g_2=g_3=g_4$ still
leads to the same universality class, several 
subspaces in $(g_1,g_2,g_3,g_4)$ have
not only  time-reversal symmetry but
also additional enhanced symmetry specified below, 
and therefore belong to different universality classes.

\subsection{Symmetries of the model}

In this subsection, we examine
symmetries of the present model, which plays a crucial
role in the classification of the universality classes for disordered
systems .
First of all, the most general Hamiltonian 
has unbroken time-reversal symmetry.
In the original basis, it is expressed by $\calH^\rmt=\calH$, whereas 
in the rotated basis where the Hamiltonian is described by the 
matrices in Eq. (\ref{GamAlp}), it is expressed by
\begin{eqnarray}
\calH=\calT\calH^\rmt\calT^{-1}, \quad \calT=1\otimes\sigma_1 .
\label{TimRev}
\end{eqnarray}
If the model does not have any other symmetries,
the universality class is $A$II and hence all states are localized.  
There are several subspaces in $(g_1,g_2,g_3,g_4)$, however,
on which symmetry is enhanced. To be concrete,
the Hamiltonian (\ref{ConPurHam}) has following symmetry
\begin{eqnarray}
\calH=-\calP_j\calH\calP_j^{-1},\quad
\mbox{with}\quad
\left\{
\begin{array}{cc}
\calP_1=\sigma_3\otimes\sigma_3, & \mbox{for} \quad g_3=g_4=0 \\
\calP_2=\sigma_1\otimes\sigma_1, & \mbox{for} \quad g_2=g_4=0 \\
\calP_3=\sigma_1\otimes i\sigma_2, & \mbox{for} \quad g_1=g_4=0
\end{array}
\right. .
\label{ChiSym}
\end{eqnarray}

Recently, Hatsugai et.al \cite{HWKM} proposed 
the same model as Eq. (\ref{LatHam}) but with random hoppings
instead of random on-site potential,
which we will refer to as HWKM model.
They have shown that such Hamiltonian 
has not only time-reversal symmetry but also another one
often called chiral symmetry.
It has been shown \cite{HWKM,Fuk,GLL,FabCas} 
that this symmetry is responsible for the existence of 
delocalized critical point at the zero energy.
Chiral symmetry in this case is expressed as
$\calH=-\calP_4\calH\calP_4^{-1}$ with $\calP_4=\sigma_3\otimes1$,
similarly to Eq. (\ref{ChiSym}), and the universality class is
specified as $BD$I or in other words
U$(2n)/$Sp($n$) (Gl(2$n$,R)/O(2$n$)) 
nonlinear sigma model for fermionic (bosonic) replica method.
Considering these, it is easily guessed that the symmetry (\ref{ChiSym})
also plays a crucial role in Anderson localization
for the present model.
In what follows, we refer to the symmetry in Eq. (\ref{ChiSym}) also as 
chiral symmetry.

\section{Replica method}
\label{s:RepMet}

It is so tedious to repeat similar calculations that we 
concentrate on the case $\calP_1$,
and hence we often denote $\calP_1$ just as $\calP$
for short. But with slight modification, any $\calP_j$ can
apply to the formulation. 
At the moment, the difference between $\calP_1$ and the others is
whether $\calP_j$ commutes or anticommutes with $\calT$ and
whether it is real symmetric or antisymmetric.
Surprisingly, this difference, which seems trivial at first glance,
leads to a different universality class.

In order to study the trasport properties of the model
near the zero energy,
we need to compute the Green functions 
\begin{eqnarray}
G(\pm i\epsilon)=\frac{1}{\pm i\epsilon-\calH} .
\nonumber
\end{eqnarray}
With the $\calP$ symmetry, however, we do not need to distinguish
the retarded and advanced Green functions, since they are related
with each other via the relation \cite{BCSZs}, 
\begin{eqnarray}
G(+i\epsilon)
=\calP\frac{1}{i\epsilon+\calH}\calP^{-1}
=-\calP G(-i\epsilon)\calP^{-1}.
\nonumber
\end{eqnarray}
The Green function $G(i\epsilon)$ can be computed by a replicated functional
\begin{eqnarray}
\calZ=\int\calD\psi\calD\psi^\dagger e^{-\int d^2x \calL} ,
\nonumber
\end{eqnarray}
with
\begin{eqnarray}
\calL
=\sum_{\alpha=1}^n\sum_{j,k=1}^4
 \psi_{j\alpha}^*(i\epsilon1_4-\calH)_{jk}\psi_{k\alpha}
\equiv\tr_{W_\rmr}\psi^\dagger(i\epsilon1_4-\calH)\psi,
\label{LagBefAve}
\end{eqnarray}
where the subscript $\alpha$ denotes the $n$ species in the replica space
$W_\rmr$, $\psi^\dagger_{\alpha j}=\psi^*_{j\alpha}$, 
and the usual convention of the matrix product is 
applied to the fermion matrix field $\psi^*_{j\alpha}$ 
and $\psi_{j\alpha}$.
Although such a notational convention is not always necessary,
it is quite convenient for us to save many complicated subscripts.
It should be noted that the fields $\psi^*_{\alpha j}$ and
$\psi_{\alpha j}$ are completely independent Grassmann variables.
Therefore, they obey in general different transformation laws.

\subsection{Auxiliary space}

Based on the Lagrangian (\ref{LagBefAve}), we will construct an effective
theory of order parameter fields for the replicas by averaging-over disorder.
In this process, it is quite important to fully incorporate the 
symmetries (\ref{TimRev}) and (\ref{ChiSym}) in the space $V$.
To this end, we introduce an auxiliary space
to reflect these into the replicas, 
according to the method developed by Zirnbauer \cite{Zir}.
Namely, we extend the replica space as 
$W_\rmr\rightarrow W\equiv W_\rmr\otimes W_\rmt\otimes W_\rmc$,
where $W_\rmt$ $(W_\rmc)$ stands for the space associtaed with time-reversal 
(chiral) symmetry.
We then introduce a 
$4n\times4~(4\times4n)$ matrix field $\tPsi_{\alpha j}~(\Psi_{j\alpha})$,
\begin{eqnarray}
\tPsi=
\left(
\begin{array}{c}
\tchi_+\\
\tchi_-
\end{array}
\right)
\quad
\Psi=(\chi_+,\chi_-);
\qquad
\tchi_\pm=\tchi\calP_\mp,
\quad
\chi_\pm=\calP_\pm\chi,
\label{DefPsi}
\end{eqnarray}
where
\begin{eqnarray}
\tchi=\frac{1}{\sqrt{2}}
\left(
\begin{array}{c}
\psi^\dagger \\
-\psi^\rmt\calT
\end{array}
\right),
\quad
\chi=\frac{1}{\sqrt{2}}
(\psi,\calT\psi^*),
\label{DefChi}
\end{eqnarray}
$\calP_\pm=\frac{1\pm\calP}{2}$,
and $\psi^\rmt$ stands for the transpose of $\psi$.
Since the components of these fields are not independent of each other, 
the fields are subject to the constraints
\begin{eqnarray}
\begin{array}{lll}
\tPsi=\gamma\Psi^\rmt\calT, 
&
\Psi=-\calT\tPsi^\rmt\gamma^{-1},  
&
\gamma=1_n\otimes i\sigma_2\otimes1_2 ,
\\
\tPsi=-\pi\tPsi\calP, 
&
\Psi= \calP\Psi\pi,
&
\pi=1_n\otimes1_2\otimes\sigma_3 .
\end{array} 
\label{RelPsi}
\end{eqnarray}
Note that the matrices $\gamma$ and $\pi$ are defined in $W$,
whereas $\calT$ and $\calP$ are defined in $V$.
We immediately find
\begin{eqnarray}
\Psi\tPsi=\calT(\Psi\tPsi)^\rmt\calT^{-1},\quad
\Psi\tPsi=-\calP(\Psi\tPsi)\calP^{-1}.
\label{PsitPsi}
\end{eqnarray}
These relations tell that 
the matrix $\Psi\tPsi$ in $V$ has the same symmetries as $\calH$ has. 
It is also easy to find alternative relations for the matrix
$\tPsi\Psi$ in $W$,
\begin{eqnarray}
\tPsi\Psi=\gamma(\tPsi\Psi)^\rmt\gamma^{-1},\quad
\tPsi\Psi=-\pi(\tPsi\Psi)\pi^{-1}  .
\label{wPsiPsi}
\end{eqnarray}
These relations actually specify the symmetry of the model,
as we shall see lator.

Although the potentials $\alpha_1$ and $\alpha_2$ have
time-reversal and chiral symmetries,
the converse is not true:
The symmetries (\ref{TimRev}) and (\ref{ChiSym}) admit other potentials
$\beta_1=\sigma_1\otimes1,\beta_2=\sigma_2\otimes\sigma_3,
\beta_3=1\otimes\sigma_1$ and $\beta_4=\sigma_3\otimes\sigma_1$.
The reason why there are no beta matrices in Eq. (\ref{ConDisHam}) is that 
the basic lattice model (\ref{LatHam}) includes only on-site random potential.
Namely, if we take all kinds of possible disorder for the lattice 
Hamiltonian including random 
(nearest neighbor and diagonal) hoppings
as well as random on-site potentials,
we have a continuum Hamiltonian with all kinds of possible 
disorder potentials
specified by the relations (\ref{TimRev}) and (\ref{ChiSym}).
Such a model may be referred to as a model with maximum entropy.
Since we are interested in the model with on-site random potentials,
we have to remove these $\beta$
matrices. For this purpose, we take account of additional conditions 
\begin{eqnarray}
\calH_\rmd=-\calT \calH_\rmd\calT,\quad
\calH_\rmd=\calE \calH_\rmd\calE,\quad \calE\equiv\sigma_3\otimes1 .
\label{AdiCon}
\end{eqnarray}
It is readily seen that
the potentials including beta matrices do not satisfy these relations.
Let us then introduce the fields
\begin{eqnarray}
\tcPsi=\frac{1}{2}
\left(
\begin{array}{c}
\tPsi\\
\tPsi\calE\\
-\tPsi\calT\\
-\tPsi\calE\calT
\end{array}
\right),
\quad
\cPsi=\frac{1}{2}
(\Psi,\calE\Psi,\calT\Psi,\calT\calE\Psi) .
\label{CirPsi}
\end{eqnarray}
We refer to this extra space as $W^\circ$.
It is easy to verfy that the matrix $\Psi^\circ\tPsi^\circ$
obeys the same relations that $\calH_\rmd$ does in (\ref{AdiCon}) 
and in (\ref{PsitPsi}) with extension 
$\gamma\rightarrow\gamma\otimes1_2\otimes\sigma_3$ and
$\pi\rightarrow\pi\otimes1_2\otimes\sigma_3$,
which means that the undesirable potentials are in fact excluded.
By using these fields we can reexpress the Lagrangian 
(\ref{LagBefAve}) as
\begin{eqnarray}
\calL=\tr_W(i\epsilon\omega\tPsi\Psi
-\tPsi i\slsh{\partial}\Psi)
-\tr_{W\otimes W^\circ}\tPsi^\circ\calH_\rmd\Psi^\circ,
\label{LagReExp}
\end{eqnarray}
where $\omega\equiv1_n\otimes1_2\otimes\sigma_1$ and 
$\slsh{\partial}\equiv\gamma_\mu\partial_\mu$.

\subsection{Effective Lagrangian averaged-over disorder}

Now we are ready to ensemble-average the Lagrangian (\ref{LagReExp}) 
over disorder. 
In this subsection, we assume $g_1=g_2=g$ for simplicity. 
More general case $g_1\ne g_2$ is briefly 
discussed in Sec. \ref{s:AniEff}.
The probability distribution (\ref{ProDis}) can be expressed as
\begin{eqnarray}
P[\calH_\rmd]\propto e^{-\frac{1}{2g}\tr_V\calH_\rmd^2} .
\label{ProDis2}
\end{eqnarray}
Therefore, the Lagrangian is converted into
\begin{eqnarray}
\calL
&=&\frac{1}{2g}\tr_V\calH_\rmd^2+
\tr_W(i\epsilon\omega\tPsi\Psi
-\tPsi i\slsh{\partial}\Psi)
-\tr_{W\otimes W^\circ}\tPsi^\circ\calH_\rmd\Psi^\circ,
\nonumber\\
&=&
\frac{1}{2g}\tr_V\left(\calH_\rmd+g\Psi^\circ\tPsi^\circ\right)^2
-\frac{g}{2}\tr_V(\Psi^\circ\tPsi^\circ)^2+
\tr_W(i\epsilon\omega\tPsi\Psi
-\tPsi i\slsh{\partial}\Psi),
\nonumber
\end{eqnarray}
where we have used a relation 
$\tr_{W\otimes W^\circ}\tPsi^\circ\calH_\rmd\Psi^\circ
=-\tr_V\calH_\rmd\Psi^\circ\tPsi^\circ$.
It should be noted that integration over $\calH_\rmd$ is 
readily performed, since the matrix $\Psi^\circ\tPsi^\circ$
has completely the same symmetries as $\calH_\rmd$ has. 
This is the reason why we have introduced several auxiliary spaces 
in the last subsection.
By using
$\tr_V(\Psi^\circ\tPsi^\circ)^2
=-\tr_{W\otimes W^\circ}(\tPsi^\circ\Psi^\circ)^2$ again and
the definition of (\ref{CirPsi}),
the gaussian integration over $\calH_\rmd$ leads to
\begin{eqnarray}
\calL=
\tr_W(i\epsilon\omega\tPsi\Psi
-\tPsi i\slsh{\partial}\Psi)
+\frac{g'}{2}\sum_{\tau=1}^4
\tr_W
(\tPsi\calO_\tau\Psi)^2,
\label{LagAftDis}
\end{eqnarray}
where $\calO_1=1, \calO_2=\calE, \calO_3=i\calT$,
$\calO_4=i\calT\calE$, and $g'=\frac{g}{4}$.
We mention in passing that 
if the model have maximun entropy, there appears only one 
interaction term with $\calO_1=1$ and $g'=g$.
Eq. (\ref{LagAftDis}) is an effective Lagrangian which generates the 
emsanble-averaged Green functions for the present model.
It is easy to confirm that new matrices are subject to
\begin{eqnarray}
\tPsi\calO_\tau\Psi
=\gamma(\tPsi\calO_\tau\Psi)^\rmt\gamma^{-1},
\quad
\tPsi\calO_\tau\Psi
=\mp\pi(\tPsi\calO_\tau\Psi)\pi^{-1}
\quad\mbox{for}\quad
\left\{
\begin{array}{c}
\tau=1,2 \\
\tau=3,4 ,
\end{array}
\right.
\label{wPsiOPsi}
\end{eqnarray}

So far we have obtained an effective action including four fermi
interactions by averaging over disorder. 
The next task is to introduce auxiliary (order-parameter)
fields to decouple them.
To this end, we add the following terms of
auxiliary fields $Q_\tau$ $(\tau=1,...,4)$ to 
the Lagrangian (\ref{LagAftDis}):
\begin{eqnarray}
\calL'=-\frac{1}{2g'}\sum_{\tau=1}^4\tr_W
(Q_\tau+g'\tPsi\calO_\tau\Psi+\delta_{\tau1}i\epsilon\omega)^2 .
\label{IntAuxFie}
\end{eqnarray}
Then, the Lagrangian becomes
\begin{eqnarray}
\calL=-\tr_W\tPsi i\slsh{\partial}\Psi
-\frac{1}{2g'}\sum_\tau\tr_W
\left(Q_\tau^2+i\epsilon\omega Q_1\delta_{\tau1}\right)
-\sum_{\tau}\tr_WQ_\tau\tPsi\calO_\tau\Psi .
\label{LagOrdPar}
\end{eqnarray}
Eq. (\ref{IntAuxFie}) requires that $Q_\tau$ should obey the same 
symmetrirs that $\tPsi\calO_\tau\Psi$ does in 
Eq. (\ref{wPsiOPsi}).
Therefore, 
\begin{eqnarray}
Q_\tau^\dagger=-Q_\tau,
\quad
Q_\tau
=\gamma Q_\tau^\rmt\gamma^{-1},
\quad
Q_\tau 
=\mp\pi Q_\tau\pi^{-1}
\quad\mbox{for}\quad
\left\{
\begin{array}{c}
\tau=1,2 \\
\tau=3,4 .
\end{array}
\right.
\label{ConQ}
\end{eqnarray}
The antihermiticity ensures that the integration over $Q_\tau$ in 
Eq. (\ref{LagOrdPar}) converges.

This Lgrangian (\ref{LagOrdPar}) has a global \SpSp symmetry. 
To see this, let us onsider a transformation
$\Psi\rightarrow\Psi g^{-1}$, 
$\tPsi\rightarrow g\tPsi$, and
$Q_\tau\rightarrow gQ_\tau g^{-1}$.
Since this transformation should keep the relations (\ref{ConQ}), 
$g$ is subject to 
$g^\dagger g=1, g^\rmt\gamma g=\gamma$ and $g^{-1}\pi g=\pi$.
Therefore, $g$ has the following form
\begin{eqnarray}
g=
\left(
\begin{array}{cc}
g_+&   \\
   &g_-
\end{array}
\right),
\quad
g_\pm g_\pm^\dagger=1,\quad
g_\pm\gamma_0g_\pm^\rmt=\gamma_0,
\quad \gamma_0\equiv1_n\otimes i\sigma_2 ,
\nonumber
\end{eqnarray}
where $2n\times2n$ matrices $g_\pm$ belongs to 
$\Wrt=W_\rmr\otimes W_\rmt$ and the explicit 
matrix above represent the space $W_\rmc$.
The conditions for $g_\pm$ require $g_\pm\in$ Sp($n$).

Next let us solve Eq. (\ref{ConQ}),
\begin{eqnarray}
&&
Q_\tau=
\left(
\begin{array}{cc}
 &M_\tau\\
-M_\tau^\dagger& 
\end{array}
\right),
\quad \mbox{for}\quad \tau=1,2,
\nonumber\\
&&
Q_\tau=
\left(
\begin{array}{cc}
A_{\tau-2}^+&\\
&A_{\tau-2}^- 
\end{array}
\right),
\quad \mbox{for}\quad \tau=3,4 ,
\nonumber
\end{eqnarray}
where $2n\times 2n$ matrices $M_\tau$ and $A_\tau^\pm$
are subject to 
$M_\tau=-\gamma_0M_\tau^*\gamma_0^{-1}$,
$A_\tau^\pm=\gamma_0 A_\tau^{\pm\rmt}\gamma_0^{-1}$,
and $A_\tau^{\pm\dagger}=-A_\tau^\pm$.
Using the definition in Eq. (\ref{DefPsi}), 
the Lagrangian is now written as
\begin{eqnarray}
\calL=&&
-\tr_\Wrt
 (\tchi_+ i\slsh{\partial}\chi_+
 +\tchi_- i\slsh{\partial}\chi_-)
+\frac{1}{2g'}\sum_{\tau=1}^2\tr_\Wrt
\left[
 2M_\tau M_\tau^\dagger -(A_\tau^{+2}+ A_\tau^{-2})
\right]
-\frac{i\epsilon}{2g'}\tr_\Wrt
(M_1-M_1^\dagger)
\nonumber\\
&&
-\sum_{\tau=1}^2\tr_\Wrt
\left(
  M_\tau\tchi_-\calO_\tau\chi_+ 
 -M_\tau^\dagger\tchi_+\calO_\tau\chi_-
 +A_\tau^+\tchi_+\calO_{\tau+2}\chi_+ 
 +A_\tau^-\tchi_-\calO_{\tau+2}\chi_-
\right) .
\label{ChiLag}
\end{eqnarray}
The tranformation laws under $g$ are
\begin{eqnarray}
\chi_\pm\rightarrow\chi_\pm g_\pm^\dagger,\quad
\tchi_\pm\rightarrow g_\pm\tchi_\pm,\quad
M_\tau\rightarrow g_+M_\tau g_-^\dagger,\quad
A_\tau^\pm \rightarrow g_\pm A_\tau^\pm  g_\pm^\dagger .
\label{TraLaw}
\end{eqnarray}

So far we have examined the case $\calP_1$. It has been shown that
the Lagrangian has \SpSp symmetry and hence the model belongs to $C$I,
which is the same class as $d$-wave superconductors with unbroken 
time-reversal and spin-rotation symmetries.
Contrary to this, in the other cases associated with 
$\calP_2$ and $\calP_3$, symmetry of the Lagrangian is quite different.
Similar calculations above show that these Lagrangians have U($2n$) symmetry 
insted of \SpSp dut to the fact that $\gamma$ defined in (\ref{RelPsi})
is $\gamma=1_n\otimes i\sigma_2\otimes\sigma_1$ if we take the same $\pi$.
This is the same class as HKWM model, which is often
referred to as $BD$I or chiral orthogonal class.
\subsection{Effective potential}
\label{s:EffPot}

Based on the Lagrangian in the last subsection, we derive the 
effective potential to study the spontaneous symmetry breaking. 
We assume that the order-paramters are diagonal
in the space $W_\rmr$. Namely, set
\begin{eqnarray}
M_\tau=
i~\mbox{diag}(m_{\tau1},...,m_{\tau n})\otimes1_2,
\quad
A_\tau^\pm=
i~\mbox{diag}(a_{\tau1}\pm b_{\tau1},...,a_{\tau n}\pm b_{\tau n})
\otimes1_2,
\quad \tau=1,2,
\nonumber
\end{eqnarray}
where $m_{\tau \alpha}(x), a_{\tau \alpha}(x)$ and 
$b_{\tau \alpha}(x)$ are real fields.
Then the Lagrangian (\ref{ChiLag}) becomes
\begin{eqnarray}
\calL=
\frac{2}{g'}\sum_{\tau=1}^2\sum_{\alpha=1}^n
\left(
m_{\tau\alpha}^2+a_{\tau\alpha}^2+b_{\tau\alpha}^2
\right)
+\frac{2\epsilon}{g'}\sum_{\alpha=1}^n m_{1\alpha}
+\sum_{\alpha=1}^n
\psi_\alpha^\dagger
\left(-i\slsh{\partial}+v_\alpha\right)\psi_\alpha,
\nonumber
\end{eqnarray}
where $v_\alpha$ is $4\times4$ matrix defined by
\begin{eqnarray}
v_\alpha=
-i (m_{1\alpha}1\otimes1+m_{2\alpha}\sigma_3\otimes1)
+a_{1\alpha}1\otimes\sigma_1
-ib_{1\alpha}\sigma_3\otimes\sigma_2
+a_{2\alpha}\sigma_3\otimes\sigma_1
-ib_{1\alpha}1\otimes\sigma_2 .
\nonumber
\end{eqnarray}
Integrating out the fermi fields, we have an effective potential 
$V_\alpha(\{m,a,b\})=S(\{m(x)=m,a(x)=a,b(x)=b\})/\mbox{volume}$
of one-loop order,
\begin{eqnarray}
V_\alpha(\{m,a,b\})=
\frac{2}{g'}\sum_{\tau=1}^2
\left(
m_{\tau\alpha}^2+a_{\tau\alpha}^2+b_{\tau\alpha}^2
\right)
-\int\frac{d^2k}{(2\pi)^2}\ln{\det}_V(\slsh{k}+v_\alpha)
+\frac{2\epsilon}{g'}\sum_{\alpha=1}^n m_{1\alpha} .
\nonumber
\end{eqnarray}
Since the potential $V_\alpha$ is a function of 
six variables, it is hard to find out the true minimum. 
However, it is readily expected from the symmetry breaking term 
that $m_{1\alpha}$ has a finite value.
Assuming it, we have tried to find out nontrivial sulution 
$\partial V_{\alpha}/\partial m_{\tau\alpha}=
 \partial V_{\alpha}/\partial a_{\tau\alpha}=
 \partial V_{\alpha}/\partial b_{\tau\alpha}=0$, 
but in vain.
Although possibility of some other solutions still remains, 
the solution $m_{1\alpha}\ne0$ and others=0 is of course 
a natural solution.
In this case, the saddle point equaiton for $m_{1\alpha}$ becomes
\begin{eqnarray}
0=\frac{\partial V_\alpha}{\partial m_{1\alpha}}=
\frac{4m_{1\alpha}}{g'}
-2\int\frac{d^2k}{(2\pi)^2}\frac{2m_{1\alpha}}{k^2+m_{1\alpha}^2},
\nonumber
\end{eqnarray}
which gives nontrivial solution
$m_{1\alpha}=m\equiv\Lambda(e^{\frac{4\pi}{g'}}-1)^{\frac{1}{2}}\sim
\Lambda e^{-\frac{2\pi}{g'}}$.
This result suggests that the operator $(\tPsi\calO_1\Psi)^2$
is relevant whereas others are irrelaevant around the present saddle-point. 
Therefore, in the following analysis, 
we neglect the irrelavant interactions and 
take only the relevant interaction into account.
In the next section, 
we will discuss slow fluctuation aroung the saddle point. 
But before it, we will see how the anisotropy $g_1\ne g_2$ modifies 
the theory so far obtained.

\subsection{Anisotropy effects}
\label{s:AniEff}

In the case where $g_1=g_2$, we could treat them simultaneously in 
Eq. (\ref{ProDis2}). Contrary to this simple case, 
we have to treat the two potentials $\alpha_1$ and $\alpha_2$
separately in anisotropic case. To this end, first introduce a new
matrix $\calF=\sigma_1\otimes1_2$ under which they transform
\begin{eqnarray}
\alpha_1=\calF\alpha_1\calF^{-1},\quad
\alpha_2=-\calF\alpha_2\calF^{-1}.
\nonumber
\end{eqnarray}
We next introduce 
\begin{eqnarray}
\tcPsi_1=\frac{1}{\sqrt{2}}
\left(
\begin{array}{l}
\tcPsi \\
\tcPsi\calF
\end{array}
\right),
\quad
\tcPsi_2=\frac{1}{\sqrt{2}}
\left(
\begin{array}{l}
\tcPsi\\
-\tcPsi\calF
\end{array}
\right), \quad
\cPsi_1=\cPsi_2=\frac{1}{\sqrt{2}}(\cPsi,\calF\cPsi).
\nonumber
\end{eqnarray}
Let us denote $\calH_{\rmd j}\equiv v_j\alpha_j$ for $j=1,2$ and
the extra auxiliary space as $W'$. 
Since the matrix $\cPsi_j\tcPsi_j~(j=1,2)$ has completely the same 
symmetry property as $\calH_{\rmd j}$, 
we can average over disorder in the following way:
\begin{eqnarray}
\frac{1}{2g_1}\tr_V\calH_{\rmd1}^2+\frac{1}{2g_2}\tr_V\calH_{\rmd2}^2
&&
-\tr_{W\otimes W^\circ\otimes W'}
\left(\tcPsi_1\calH_{\rmd1}\cPsi_1+\tcPsi_2\calH_{\rmd2}\cPsi_2\right)
\nonumber\\
&&\rightarrow
\frac{g_1}{2}\tr_{W\otimes W^\circ\otimes W'}
(\tcPsi_1\cPsi_1)^2+\frac{g_2}{2}\tr_{W\otimes W^\circ\otimes W'}
(\tcPsi_1\cPsi_1)^2
\nonumber\\
&&=
\frac{g_1+g_2}{2}\tr_{W\otimes W^\circ}
(\tcPsi\cPsi)^2
+\frac{g_1-g_2}{2}\tr_{W\otimes W^\circ}
(\tcPsi\calF\cPsi)^2 ,
\nonumber
\end{eqnarray}
where arrow menas that the integration over 
$\calH_{\rmd j}$ is carried out.
This equaiton shows that 
anisotropy $g_1\ne g_2$ yields an additional term including $\calF$.
However, it turns out that this term is also irrelevant and therefore
can be neglected, following similar discussions to the previous subsection. 

\section{Effective action for the Goldstone mode}
\label{s:ActGol}

The saddle-point $M_1=m 1_n\otimes1_2$
is invariant under $g_+=g_-$ type transformation in Eq. (\ref{TraLaw}) but 
not invarinat under $g_+=g_-^\dagger$.
This means that \SpSp symmetry is broken to Sp($n$). 
Therefore, fluctuation around the saddle-point can be incorporated
by parametrizing the order-parameter field $M_1$ as
\begin{eqnarray}
M_1=\xi im \xi=im \xi^2=im U,
\nonumber
\end{eqnarray}
where $\xi$ and $U\in$ Sp($n$), i.e., 
$U^\rmt\gamma_0U=\gamma_0$, $U^\dagger U=1_{2n}$.
The field $U$ descirbes the Goldstone mode
around the saddle point.
Then, the Lagrangian becomes
\begin{eqnarray}
\calL=-\tr_\Wrt
\left[
\tchi_+i\slsh{\partial}\chi_+ +
\tchi_-i\slsh{\partial}\chi_- +
im(U\tchi_-\chi_++U^\dagger\tchi_+\chi_-)
\right]
+\frac{\epsilon m}{2g'}\tr_\Wrt(U+U^\dagger)+\frac{2nm^2}{g'}
\nonumber
\end{eqnarray}
The last constant term will be neglected, since it vanishes in the
replica limit $n\rightarrow0$.
In the following subsection, 
we construct an effective theory for small fluctuation
$U$ around the saddle point.

\subsection{Integration over fermi fields}

So far we have treated $\tchi$ and $\chi$ as matrix fields to save 
complicated subscripts. 
It has actually worked well especially for the discussions of the symmetry. 
However, in order to extract an effective 
Lagrangian for $U$ via chiral gauge transformation introduced below, 
we cannot treat the space $V$ and $W$ separately any longer. 
Therefore, we switch into the standard notation for fermi fields as follows:
Note the following identity, for example, 
$
\tr_\Wrt U\tchi_-\chi_+
=
\sum_{\alpha,\alpha'}\sum_{j}
U_{\alpha'\alpha}\tchi_{-\alpha j}\chi_{+j\alpha'}
=
\sum_{j,\alpha}\sum_{j',\alpha'}
\tchi_{-\alpha j}\delta_{jj'}
\otimes U_{1\alpha\alpha'}^\rmt\chi_{+j'\alpha'}
=
\sum_{j,\alpha}\sum_{j',\alpha'}
\tchi_{-\alpha j}\delta_{jj'}
\otimes \gamma_0U_{1\alpha\alpha'}^\dagger\gamma_0^{-1}\chi_{+j'\alpha'}
\rightarrow
\tchi_- 1_4\otimes U^\dagger\chi_+
$,
where the last arrow means that we have made a $\gamma_0$ rotation to 
$\tchi$ and $\chi$ and we have regarded $\chi_\pm$ as a vector with
$8n$ components in $V\otimes\Wrt$.
Then the fermion part of the Lagrangian can be denoted as
\begin{eqnarray}
-\calL_F&=&
\tchi_+i\slsh{\partial}\otimes1_{2n}\chi_+ +
\tchi_-i\slsh{\partial}\otimes1_{2n}\chi_- +
im\left(
\tchi_- 1_4\otimes U^\dagger\chi_+ +
\tchi_+ 1_4\otimes U\chi_-
\right)
\nonumber\\
&=&
\tchi
\left[
i\slsh{\partial}\otimes1_{2n}+
im
\left(
\calP_+\otimes U^\dagger+\calP_-\otimes U
\right)
\right]
\chi .
\label{ChiStaLag}
\end{eqnarray}
Although this Lagrangian seems to be standard, the integration 
over $\chi$ is not straightforward. 
It is because the elementary fields $\psi^\dagger$
and $\psi$ are doubly included in $\tchi$ and $\chi$ defined 
in Eq. (\ref{DefChi}).
It is possible, however, to rewrite this Lagrangian as
\begin{eqnarray}
-\calL_F=
\tpsi'
\left(
\begin{array}{cc}
imU^\dagger                     & i(\partial_2-i\partial_1)1_{2n} \\
i(\partial_2+i\partial_1)1_{2n} & imU
\end{array}
\right)
\psi' ;\qquad
\tpsi'=
(\psi_1^\dagger,-\psi_3^\rmt,\psi_2^\dagger,-\psi_4^\rmt),
\quad
\psi'=
\left(
\begin{array}{l}
\psi_1 \\ \psi_3^* \\ \psi_2 \\ \psi_4^*
\end{array}
\right) ,
\label{SimLag}
\end{eqnarray}
where $\tpsi'$ and $\psi'$ have $4n$ components whose 
explicit subscripts in the above definition are those for $V$.
It should be noted that this Lagrangian is described by 
independent field variables $\tpsi'$ and $\psi'$ only.
As a result, we can integrate out fermi fields, which results in
\begin{eqnarray}
\int\calD\psi\calD\psi^\dagger e^{\int d^2x\calL_F}
&=&\Det_{V\otimes\Wr}
\left(
\begin{array}{cc}
imU^\dagger                     & i(\partial_2-i\partial_1)1_{2n} \\
i(\partial_2+i\partial_1)1_{2n} & imU
\end{array}
\right)
\nonumber\\
&=&\Det_{V\otimes\Wrt}^{\frac{1}{2}}
\left[
i\slsh{\partial}\otimes1_{2n}+
im\left(
\calP_+\otimes U^\dagger+\calP_-\otimes U
\right)
\right] .
\label{FerDet}
\end{eqnarray}
Based on this result, we derive an effective action for the
fluctuation $U$ in the next subsection.

\subsection{Chiral gauge transformation}

Suppose that the fluctuation $U(x)$ around the saddle-point 
varies slowly with $x$.
Then, we can expand, with respect to derivatives,
the determinant derived in the last subsection.
To this end, make the folowing transformation 
\begin{eqnarray}
\begin{array}{ll}
\tchi_+^U= \tchi_+U, \quad& 
\chi_+^U= U^\dagger \chi_+, \\
\tchi_-^U=\tchi_-,\quad &\chi_-^U=\chi_-.
\end{array}
\label{ChiTra}
\end{eqnarray}
Here and hereafter, we often
denote $1_4\otimes U$ just as $U$ for simplicity.
Without confusion, 
we also use e.g., $\calP$ instead of $\calP\otimes1_{2n}$. 
The Lagrangian is now converted into
\begin{eqnarray}
-\calL_F=\tchi^U(i\slsh{\hat{D}}+im)\chi^U,
\nonumber
\end{eqnarray}
where
\begin{eqnarray}
\hat{D}_\mu=D_{+\mu}+\partial_{-\mu};
\qquad
D_{+\mu}=D_\mu\calP_+,
\quad
\partial_{-\mu}=\partial_{\mu}\calP_-  ,
\label{DirOpe}
\end{eqnarray}
with
\begin{eqnarray}
D_\mu=\partial_\mu+A_\mu, \quad 
A_\mu=U^\dagger\partial_\mu U .
\nonumber
\end{eqnarray}
Integration over fermi fields $\tchi^U$ and $\chi^U$ gives also a determinant,
and the action for $U$ is 
\begin{eqnarray}
S=-\frac{1}{2}\ln\Det_{V\otimes\Wrt}(i\slsh{\hat{D}}+im)
-\ln J(U)+\frac{\epsilon m}{2g'}\int d^2x\tr_{\Wrt}(U+U^\dagger)
\label{ActChiTra}
\end{eqnarray}
where $J(U)$ is a Jacobian \cite{Fuj} due to the transformation (\ref{ChiTra}).
It turns out that expansion of the determinant yields no second order
terms with respect to the derivative, and hence we neglect the
contribution from it in the leading order approximation.
In what follows, we then concnetrate on the calculation of the Jacobian.
We mention in passing that 
we will use the Lagrangian (\ref{ChiStaLag}) to compute the Jacobian,
although (\ref{SimLag}) is less confusing.

Let us consider the following eigenvalue problem for the Dirac 
opeartor (\ref{DirOpe}) to define and regularize the Jacobian of 
the path integration,
\begin{eqnarray}
i\slsh{\hat{D}}\varphi_n=\lambda_n\varphi_n,\quad
\phi_n^\dagger i\slsh{\hat{D}}=\phi_n^\dagger\lambda_n,
\nonumber
\end{eqnarray}
where right- and left-eigenvalue equations are introduced, since 
the Dirac operator is non-hermitian.
The eigenfunctions are normalized via the inner product
$(\phi_n^\dagger,\varphi_m)\equiv\int d^2x\phi_n^\dagger\varphi_m
=\delta_{nm}$.
Assuming the completeness of the eigenfunctions, the fields
$\tchi$ and $\chi$ are expanded as
\begin{eqnarray}
\begin{array}{ll}
\chi^U=\sum_na_n^U\varphi_n, \quad&
a_n^U=(\phi_n^\dagger,\chi^U) ,
\\
\tchi^U=\sum_n\phi_n^\dagger \tb_n^U, \quad&
\tb_n^U=(\tchi^U,\varphi_n)  .
\end{array} 
\nonumber
\end{eqnarray}
The measure is then defined as
\begin{eqnarray}
\calD\chi^U=\prod_n da_n^U
\quad\mbox{or}\quad
\calD\tchi^U=\prod_nd\tb_n^U.
\nonumber
\end{eqnarray}
It should be noted that if one uses the Lagrangian (\ref{SimLag}) 
described by independent field variables $\tpsi'$ and $\psi'$ only,
the measure is defined by $\calD\psi'\calD\tpsi'$ as usual. 
In the present case, however, either $\calD\chi$ or $\calD\tchi$ is
enough, since each already includes all independent variables. 
It seems to be unusual, to be sure, but we follow the
calculation with the latter notation.
Using the definitions for the path integration measure, 
we can compute the Jacobian for 
infinitesimal small change for the gauge transformation (\ref{ChiTra}),
$\chi^U \rightarrow \chi^{U+\delta U}= \chi^U+\delta \chi^U$ and
$\tchi^U\rightarrow\tchi^{U+\delta U}=\tchi^U+\delta\tchi^U$ with
$\delta\chi^U=-\calP_+\otimes U^\dagger\delta U\chi^U$ and
$\delta\tchi^U=\tchi^U\calP_-\otimes U^\dagger \delta U$
as follows \cite{Fuj}:
\begin{eqnarray}
&&
a_n^{U+\delta U}=(\phi_n^\dagger,\chi^U+\delta\chi^U)=
\sum_m(1-N)_{nm}a_m^U,
\quad
\nonumber\\
&&
\tb_n^{U+\delta U}=(\tchi^U+\delta\tchi^U,\varphi_n)
=\sum_m\tb_m^U(1-\widetilde N)_{mn},
\nonumber
\end{eqnarray}
where
\begin{eqnarray}
&&
N_{mn}=(\phi_m^\dagger,\calP_+\otimes U^\dagger\delta U\varphi_n),
\nonumber\\
&&
\widetilde N_{mn}=
-(\phi_m^\dagger\calP_-\otimes U^\dagger\delta U,\varphi_n) .
\nonumber
\end{eqnarray}
Therefore, the measure changes as 
$\prod da_n^U =\Det_{V\otimes\Wrt\otimes M}
(1-N)           \prod  da_n^{U+\delta U}$ and
$\prod d\tb_n^U=\Det_{V\otimes\Wrt\otimes M}
(1-\widetilde N)\prod d\tb_n^{U+\delta U}$, where $M$ stands for
the space for the eigenmodes of the Dirac operator.
Noting that $\det(1+\epsilon X)\sim e^{\epsilon \tr X+O(\epsilon^2)}$
valid for small $\epsilon$, we reach 
\begin{eqnarray}
\calD \chi^U=\calD \chi^{U+\delta U}e^{-\delta\Gamma_\chi},
\nonumber\\
\calD\tchi^U=\calD\tchi^{U+\delta U}e^{-\delta\Gamma_\tchi},
\nonumber
\end{eqnarray}
where 
$\delta\Gamma_\chi= \Tr_{V\otimes\Wrt\otimes M}           N
\equiv\int d^2x\sum_n\tr_{V\otimes\Wrt}N_{nn}$ and
$\delta\Gamma_\tchi=\Tr_{V\otimes\Wrt\otimes M}\widetilde N
\equiv\int d^2x\sum_n\tr_{V\otimes\Wrt}\widetilde N_{nn}$.
Here $\Tr_M$ stands for the trace for $x$ and $M$.
In principle, we can choose either $\Gamma_\chi$ or $\Gamma_\tchi$ 
as the Jacobian in Eq. (\ref{ActChiTra}) if they can be regularized. 
In what follows, we will apply standard regularization scheme in Refs. 
\cite{Fuj,AlvGin}. Nevertheless,
it turns out that they still give rise to divergent terms 
for each $\Gamma$. Fortunately, as we shall see below,
they cancel each other in $\delta\Gamma_\chi+\delta\Gamma_\tchi$. 
Therefore, it is suitable to define the Jacobian in Eq. (\ref{ChiTra})
as a square root of the product of the naive Jacobians as
$\Det_{V\otimes\Wrt\otimes M}^{\frac{1}{2}}(1-N)(1-\widetilde N)$. 
Namely,
\begin{eqnarray}
\delta\ln J(U)\equiv\delta\Gamma=
\frac{1}{2}(\delta\Gamma_\chi+\delta\Gamma_\tchi).
\nonumber
\end{eqnarray}
It is possible to show that the Jacobian based on 
$\calD\psi'\calD\tpsi'$ for the Lagrangian (\ref{SimLag}) indeed
yields the same Jacobian.

We now give the outline of how to calculate the Jacobian
following Refs. \cite{Fuj,AlvGin}.
Using the definition of the $\Tr_M$, we have 
\begin{eqnarray}
\delta\Gamma
&=&
\frac{1}{2}\Tr_{V\otimes\Wrt}\sum_n
\phi_n^\dagger(x)\calP\otimes U^\dagger\delta U\varphi_n(x)
\nonumber\\
&=&
\frac{1}{2}\Tr_{V\otimes\Wrt}\sum_n
\phi_n^\dagger(x)\calP\otimes U^\dagger\delta U
e^{-\left( \lambda_n /\Lambda\right)^2}\varphi_n(x)
\nonumber\\
&\equiv&
\frac{1}{2}\Tr_{V\otimes\Wrt\otimes M}
\calP\otimes U^\dagger\delta U
e^{-\left( i \not\hat D/\Lambda\right)^2}
\nonumber
\end{eqnarray}
where ultraviolet cut-off $\Lambda$ is introduced, which will be set 
$\Lambda\rightarrow\infty$ at the end of the calculation.
Using the relation 
$\slsh{\hat D}^2=
\slsh{D}_+\slsh{\partial}_-+\slsh{\partial}_-\slsh{D}_+$,
we have
\begin{eqnarray}
\delta\Gamma
&=&
\frac{1}{2}\Tr_{V\otimes\Wrt\otimes M}U^\dagger\delta U
\left[
\calP(e^{\not \partial\not D / \Lambda^2}
+e^{\not D\not\partial/\Lambda^2})
+
(e^{\not \partial\not D/\Lambda^2}
-e^{\not D\not\partial/\Lambda^2})
\right]
\nonumber\\
&=&
\frac{1}{2}\int\frac{d^2k}{(2\pi)^2}
\Tr_{V\otimes\Wrt}e^{-ikx}U^\dagger\delta U
\left[
\calP(e^{\not\partial\not D/\Lambda^2}
+e^{\not D\not\partial/\Lambda^2})
+
(e^{\not\partial\not D/\Lambda^2}
-e^{\not D\not\partial/\Lambda^2})
\right]e^{ikx},
\nonumber
\end{eqnarray}
where in the last line we have switched the basis for $M$ 
from the eigenmodes of the Dirac operator into plane waves.
Expanding $\Lambda$ and taking the limit $\Lambda\rightarrow\infty$,
we reach
\begin{eqnarray}
\delta\Gamma=
\frac{1}{2}\int\frac{d^2k}{(2\pi)^2}e^{-k^2}
\Tr_{V\otimes\Wrt}
\left[
\calP\otimes U^\dagger\delta U(i\partial\times D)+
1\otimes U^\dagger\delta U(\partial\cdot D)
\right] .
\label{DelGam}
\end{eqnarray}
where
$a\times b=\sigma_{\mu\nu}a_\mu b_\nu$ and 
$a\cdot  b=g_{\mu\nu}a_\mu b_\nu$ with
\begin{eqnarray}
\sigma_{\mu\nu}=\frac{1}{2i}[\gamma_\mu,\gamma_\nu]
=\epsilon_{\mu\nu}\sigma_3\otimes\sigma_3,
\quad
g_{\mu\nu}=\frac{1}{2}\{\gamma_\mu,\gamma_\nu\}
=\delta_{\mu\nu}1\otimes1.
\label{SigG}
\end{eqnarray}
Therefore, taking trace in $x$- and $V$-space, we have
\begin{eqnarray}
\delta\Gamma=\frac{1}{2\pi}\int d^2x\tr_\Wrt U^\dagger\delta U
\left(\delta_{\mu\nu}-i\epsilon_{\mu\nu}
\right)
\partial_\mu A_\nu .
\nonumber
\end{eqnarray}
It is nowadays well-known that the action function satisfying above
includes not only the usual kinetic term but also the WZW term,
\begin{eqnarray}
\Gamma = S_{\rmWZW}^{k} \equiv \frac{1}{4\pi}\int d^2x\tr_\Wrt
\partial_\mu U\partial_\mu U^\dagger+\frac{ik}{12\pi}
\int d^3x\epsilon_{\mu\nu\lambda}\tr_\Wrt 
U^\dagger\partial_\mu UU^\dagger\partial_\nu
UU^\dagger\partial_\lambda U ,
\label{WZWAct}
\end{eqnarray}
with $k=2$. 
Adding the symmetry-breaking term in (\ref{ActChiTra}),
we end up with an action  
\begin{eqnarray}
S=S_{\rmWZW}^{k=2}+\frac{\epsilon m}{2g'}\int
d^2x\tr_\Wrt(U+U^\dagger) .
\label{FinAct}
\end{eqnarray}
Therefore, it turns out that the Goldstone mode of the model with
chiral symmetry of type $\calP_1$
can be described by Sp($n)_2$ WZW model (\ref{WZWAct}) with a symmetry
breaking term as in (\ref{FinAct}).

So far we have considered the case where Hamiltonian has the chiral
symmetry specified by $\calP_1$ in Eq. (\ref{ChiSym}). 
As to the other cases, similar calculations lead to 
U($2n$)/Sp($n$) nonlinear sigma model, which is so-called chiral 
orthogonal class or $BD$I,  
the same universality class as HWKM model.

\section{Phase diagram}
\label{s:PhaDia}

\begin{figure}[htb] 
\epsfxsize=100mm 
\centerline{\epsfbox{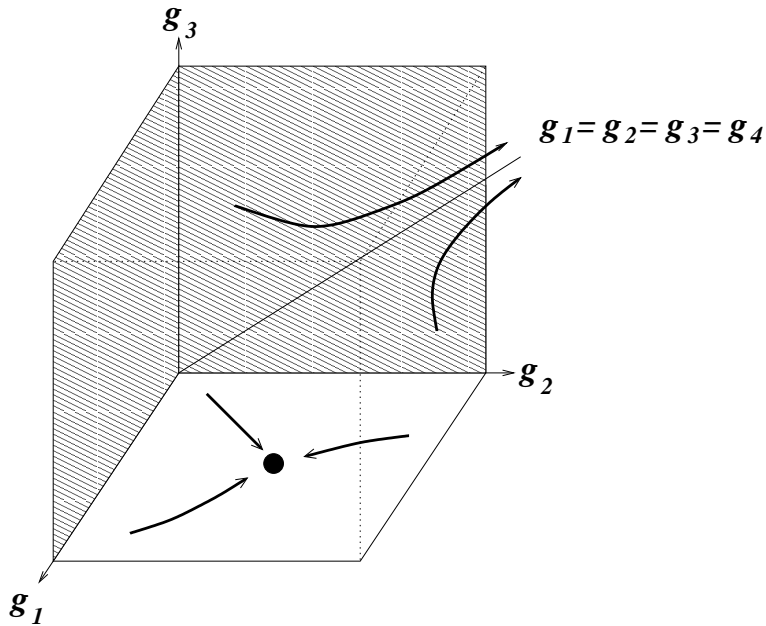}} 
\vspace{0.2cm}
\caption{Schematic phase diagram at the zero energy in the case of
weak disorder.
We omitted the axis $g_4$. Black circle indicates the Sp($n)_2$ WZW
fixed point. 
Shaded surfaces stand for ``critical surfaces'' described by U($2n$)/Sp($n$) 
nonlinear sigma model. Any points away from these three surfaces flow to 
strong coupling limit of O($2n,2n$)/O($2n)\times$O($2n$)
nonlinear sigma model.}
\label{f:fig2}
\end{figure}

We present a schematic phase diagram at the zero energy 
in Fig. \ref{f:fig2}, which is valid for weak disorder, 
i.e, for small $g_j$.
In the case $g_3=g_4=0$, which is denoted by a white surface in the figure,
the model belongs to class $C$I and 
is described by Sp($n)_2$ WZW model with some irrelevant
perturbations discussed in Sec. \ref{s:EffPot} and \ref{s:AniEff}.
The reason why the present WZW model is level-2 is that 
the continuum limit (\ref{ConPurHam}) leads to double Dirac fermions. 
One of the main origin of the irrelevant interactions is that the lattice 
model has only on-site potentials without any hopping randomness.
Namely, the model does not have maximum  entropy.
Another origin is the anisotropy $g_1\ne g_2$, as discussed in
Sec. \ref{s:AniEff}.
Besides, the continuum model should have basically many higher order  
irrelevant interactions.
Therefore, at the zero energy 
any points on this surface flow to the Sp($n)_2$ 
WZW fixed point, denoted by a black circle in the figure.
It turns out that
around the fixed point the density of states obeys
the scaling law $\rho(E)\sim E^{\frac{1}{11}}$, following from the
exact results obtained by Knizhnik and Zamolodchikov \cite{KniZam}.

On the other hand, the universality class of the two surfaces 
$g_1=g_4=0$ and $g_2=g_4=0$, shaded in the figure, is different. 
They belong to 
the class of U($2n$)/Sp($n$) nonlinear sigma model or in other words,
chiral orthogonal class or class $BD$I, 
which is the same universality class as that of HWKM model.
It is well-known that 
the beta function of the renormalization 
group vanishes and the coupling constant of the nonlinear sigma model
is not renormalized.
Hence any points on these surfaces are critical. 
The density of states is conjectured to be divergent at the zero
energy, according to 
$\rho(E)\sim\frac{\Lambda}{E}\exp[-\alpha\sqrt{\ln(\Lambda/E)}]$
with a positive constant $\alpha$ \cite{Gad}.

These surfaces, one $C$I and two $BD$I's, are unstable: Out the surfaces
renormalization group flows are toward the strong-coupling limit of 
Sp($2n$)/\SpSp nonlinear sigma model, as 
shown by Fisher and Fradkin for  $g_1=g_2=g_3=g_4$ case.
Therefore, all states away from the surfaces are expected to be localized.

\section{Summary and concluding remarks}
\label{s:Sum}

We have studied the phase diagram of the disordered fermion model
with random on-site potential defined on a two-dimensional square lattice 
with $\pi$-flux.
The continuum limit near the zero energy yields Dirac fermions
with random potentials specified by four independent coupling constants.
It has been argued that
in addition to time-reversal symmetry,
this model has so-called chiral symmetry in 
three subspaces in the space of the coupling constants.
Here, chiral symmetry means that there exist a matrix anticommuting
with the Hamiltonian.
What is interesting is that this model alows three kinds of 
chiral symmetry, which is due to the species doubling of the lattice
fermion. It has been shown that one of them leads to \SpSp symmety of the
Lagrangian for generating functional of emsemble-averaged Green functions 
whereas the others lead to U($2n$) symmetry, both of which 
is broken to Sp($n$).
It has been found that 
the fluctuation around the saddle point is described, respectively, by
Sp($n)_2$ WZW model for the former 
and U($2n$)/Sp($n$) nonlinear sigma model for the latter.
Based on these results, we have proposed a phase diagram.

The Dirac fermion model with $\calP_1$ symmetry 
calculated in detail in the text
is quite similar to that for the disordered $d$-wave superconductors
with spin-ratation and time-reversal symmetries.
Actually, effective Lagrangians for these systems have \SpSp symmetry
and hence both the models belong to the same universality class $C$I.
However, fixed-points are quite different due to the WZW term.
It was shown \cite{Fukd,ASZ} that in the $d$-wave superconductors 
the WZW terms for several species of Dirac fermions
cancel each other and the resultant theory predicts localization
of quasiparticles at the band center.  
Contrary to this, the model treated in this paper
is described by level-2 WZW model, 
which means that it has delocalized states.
The difference lies in Eq. (\ref{SigG}): 
In the present model the WZW term survives owing to 
the relation 
$\sigma_{\mu\nu}=[\gamma_\mu,\gamma_\nu]/2i=\epsilon_{\mu\nu}\calP_1$,
which is due to $\calP_1=i\gamma_1\gamma_2$. 
This is just the same definition of so-called $\gamma_5$ in the particle
physics literature, 
although the present $\gamma$ matrices are not $2\times2$ but $4\times4$.
Conversely, we can say that chiral symmetry only in this sense 
yields the chiral anomaly and hence the WZW term, although we often use 
the same words for other similar symmetry in the 
Anderson localization literature.
Considering the fact that 
the lattice model having $\calP_1$ chiral symmetry is realized in the
case $v_3'=-v_1'$ and $v_4'=-v_2'$ in Eq. (\ref{LatHam}), 
which is indeed a fine-tuned model, 
we conclude that the delocalization due to the existence of the WZW term 
seldom occures in nature.




\begin{references}
\bibitem[*]{Email} Email: fukui@mito.ipc.ibaraki.ac.jp
%
\bibitem{Gad}
R. Gade, 
Nucl. Phys. {\bf B398}, 499 (1993);
R. Gade and F. Wegner,
{\it ibid.}, {\bf B360}, 213 (1991).
%
\bibitem{Zir}
M. R. Zirnbauer,
J. Math. Phys. {\bf 37}, 4986 (1996).
%
\bibitem{AltZir}
A. Altland and M. R. Zirnbauer,
Phys. Rev. {\bf B55}, 1142 (1997).
%
\bibitem{AALR}
E. Abrahams, P. W. Anderson, D. C. Licciardello, 
and T. V. Ramarkrishnan,
Phys. Rev. Lett. {\bf 42}, 673 (1979).
%
\bibitem{Fur}
A. Furusaki, 
Phys. Rev. Lett. {\bf 82}, 604 (1999), and references therein.
%
\bibitem{AltSim}
A. Altland and B. D. Simons,
Nucl. Phys. {\bf B562}, 445 (1999).
%
\bibitem{GLL}
S. Guruswamy, A. LeClair, and A. W. W. Ludwig, 
cond-mat/9909143.
%
\bibitem{FabCas}
M. Fabrizio and C. Castellani,
Nucl. Phys. {\bf B583}, 542 (2000).
%
\bibitem{HWKM}
Y. Hatsugai, X.-G. Wen, and M. Kohmoto,
Phys. Rev. {\bf B56}, 1061 (1997);
Y. Morita and Y. Hatsugai, 
Phys. Rev. Lett. {\bf 79}, 3728 (1997);
Y. Morita and Y. Hatsugai, 
Phys. Rev. {\bf B58}, 6680 (1998).
%
\bibitem{Fuk}
T. Fukui,
Nucl. Phys. {\bf B562}, 477 (1999).
%
\bibitem{NTW}
A. A. Nersesyan, A. M. Tsvelik, and F. Wenger,
Phys. Rev. Lett. {\bf 72}, 2628 (1994).
%
\bibitem{ZHH}
K. Ziegler, M. H. Hettler, and P. J. Hirschfeld,
Phys. Rev. Lett. {\bf 77}, 3013 (1996).
%
\bibitem{SFBN}
T. Senthil, M. P. A. Fisher, L. Balents, and C. Nayak,
Phys. Rev. Lett. {\bf 81}, 4704 (1998):
T. Senthil and M. P. A. Fisher,
Phys. Rev. {\bf B60}, 6893 (1999).
%
\bibitem{BCSZd}
R. Bundschuh, C. Cassanello, D. Serban, and M. R. Zirnbauer,
Phys. Rev. {\bf B59}, 4382 (1999).
%
\bibitem{SenFis}
T. Senthil and M. P. A. Fisher,
Phys. Rev. {\bf B61}, 9690 (2000).
%
\bibitem{BSZ}
M. Bocquet, D. Serban, and M. R. Zirnbauer,
cond-mat/9910480.
%
\bibitem{Fukd}
T. Fukui,
cond-mat/0002002.
%
\bibitem{VisFis}
S. Vishveshwara and M. P. A. Fisher,
cond-mat/0003018.
%
\bibitem{FenKon}
P. Fendley and R. M. Konic,
cond-mat/0003436.
%
\bibitem{AHM}
W. A. Atkinson, P. J. Hirschfeld, and A. H. MacDonald,
cond-mat/0002333.
%
\bibitem{ZST}
J.-X. Zhu, D. N. Sheng, and C. S. Ting,
cond-mat/0005266.
%
\bibitem{AHMZ}
W. A. Atkinson, P. J. Hirschfeld, A. H. MacDonald, and K. Ziegler,
cond-mat/0005487.
%
\bibitem{ASZ}
A. Altland, B. D. Simons, and M. R. Zirnbauer,
cond-mat/0006362.
%
\bibitem{Pru}
A. M. M. Pruisken, Nucl. Phys. {\bf B235}, 277 (1984);
H. Levine, S. B. Libby, and A. M. M. Pruisken, {\it ibid} {\bf B235}, 
30, 49, 71 (1984).
%
\bibitem{FisFra}
M. P. A. Fisher and E. Fradkin,
Nucl. Phys. {\bf B251}, 457 (1985).
%
\bibitem{LFSG}
A. W. W. Ludwig, M. P. A. Fisher, R. Shankar, and G. Grinstein,
Phys. Rev. {\bf B50}, 7526 (1994).
%
\bibitem{BCSZs}
R. Bundschuh, C. Cassanello, D. Serban, and M. R. Zirnbauer,
Nucl. Phys. {\bf B532}, 689 (1998).
%
\bibitem{Fuj}
K. Fujikawa,
Phys. Rev. {\bf D21}, 2848 (1980);
{\it ibid} {\bf D22}, 1499 (1980) (E)
%
\bibitem{AlvGin}
L. Alvarez-Gaum\'e and P. Ginsparg,
Nucl. Phys. {\bf B243}, 449 (1984).
%
\bibitem{KniZam}
V. G. Knizhnik and A. B. Zamolodchikov,
Nucl. Phys. {\bf B247}, 83 (1984).
%
\end{references}
\end{document}